\documentclass[prb,reprint,superscriptaddress,floatfix,amsmath,amssymb]{revtex4-2}
\usepackage{color}
\usepackage{graphicx}
\usepackage{hyperref}
\usepackage{physics}
\usepackage[maxfloats=256]{morefloats}

\usepackage{xspace} 

\usepackage[whole]{bxcjkjatype}
\hypersetup{colorlinks=true,urlcolor=blue,citecolor=blue,linkcolor=blue,breaklinks=true}

\newcommand\sqrtthree{\ensuremath{\sqrt{3}}}
\newcommand\sqrtthreesqrtthree{\ensuremath{\sqrt{3}\times\sqrt{3}}\xspace}

\newcommand\mqzero{\ensuremath{m^2_{q=0}}\xspace}
\newcommand\fchiral{\ensuremath{\kappa^2_{\mathrm{Ferro}}}\xspace}
\newcommand\mqsqrtthree{\ensuremath{m^2_{\sqrt{3}\times\sqrt{3}}}\xspace}
\newcommand\afchiral{\ensuremath{\kappa^2_{\mathrm{AF}}}\xspace}
\newcommand\moctupole{\ensuremath{m^2_{\mathrm{oct}}}\xspace}

\newcommand\TBKT{\ensuremath{T_{\mathrm{BKT}}}\xspace}
\newcommand\Tafchiral{\ensuremath{T_{\mathrm{afchical}}}\xspace}

\definecolor{purple}{rgb}{0.6, 0.2, 0.8}
\definecolor{orange}{rgb}{0.91, 0.41, 0.17}

\begin{document}

\allowdisplaybreaks
\title{Monte Carlo study on low-temperature phase diagrams of the $J_1$-$J_2$ classical $XY$ kagome antiferromagnet}
\author{Fumiya Kakizawa}
\affiliation{Department of Physics, Saitama University, Saitama 338-8570, Japan}

\author{Takahiro Misawa}
\affiliation{Beijing Academy of Quantum Information Sciences, Haidian District, Beijing 100193, China}
\affiliation{Institute for Solid State Physics, University of Tokyo, 5-1-5 Kashiwanoha, Kashiwa, Chiba 277-8581, Japan}

\author{Hiroshi Shinaoka}
\affiliation{Department of Physics, Saitama University, Saitama 338-8570, Japan}

\begin{abstract}
Frustrated magnets with degenerate ground states exhibit exotic ground states and rich phase structures when perturbations and/or thermal fluctuations lift the degeneracy. 
In two-dimensional models with short-range interactions, continuous symmetries cannot spontaneously break at finite temperatures, leading to the suppression of conventional magnetic long-range ordering (LRO).
In this paper, we numerically study the classical $J_1$-$J_2$ $XY$ antiferromagnet on the kagome lattice as a prototype model of such frustrated magnets, where $J_2$ denotes the next-nearest-neighbor exchange interaction.
We map out the $J_2$-$T$ phase diagram of this model employing extensive classical Monte Carlo (MC) simulations.
The obtained phase diagram features Berezinskii-Kosterlitz-Thouless (BKT) transitions of $q=0$, $\sqrtthreesqrtthree$ magnetic orders, and octupole orders, in addition to finite-temperature phase transitions of both ferrochiral and antiferrochiral long-range orders. 
Additionally, we find a non-trivial first-order transition for antiferromagnetic $J_2/J_1 < 0$. 
The origin of this transition is discussed in the context of non-local loop structures present in local $120^\circ$ spin structures.
\end{abstract}

\maketitle

\section{Introduction}\label{sec:introduction}
Classical spin models on frustrated lattices, such as the triangular lattice and the kagome lattice, often have a large number of degenerate ground states at a macroscopic level. 
When the degeneracy is lifted by perturbations, such as magnetic fields, long-range interactions, and thermal fluctuations, several exotic states emerge~\cite{diep2013frustrated}. 

\begin{figure}
   \centering
   \includegraphics[width=0.8\columnwidth]{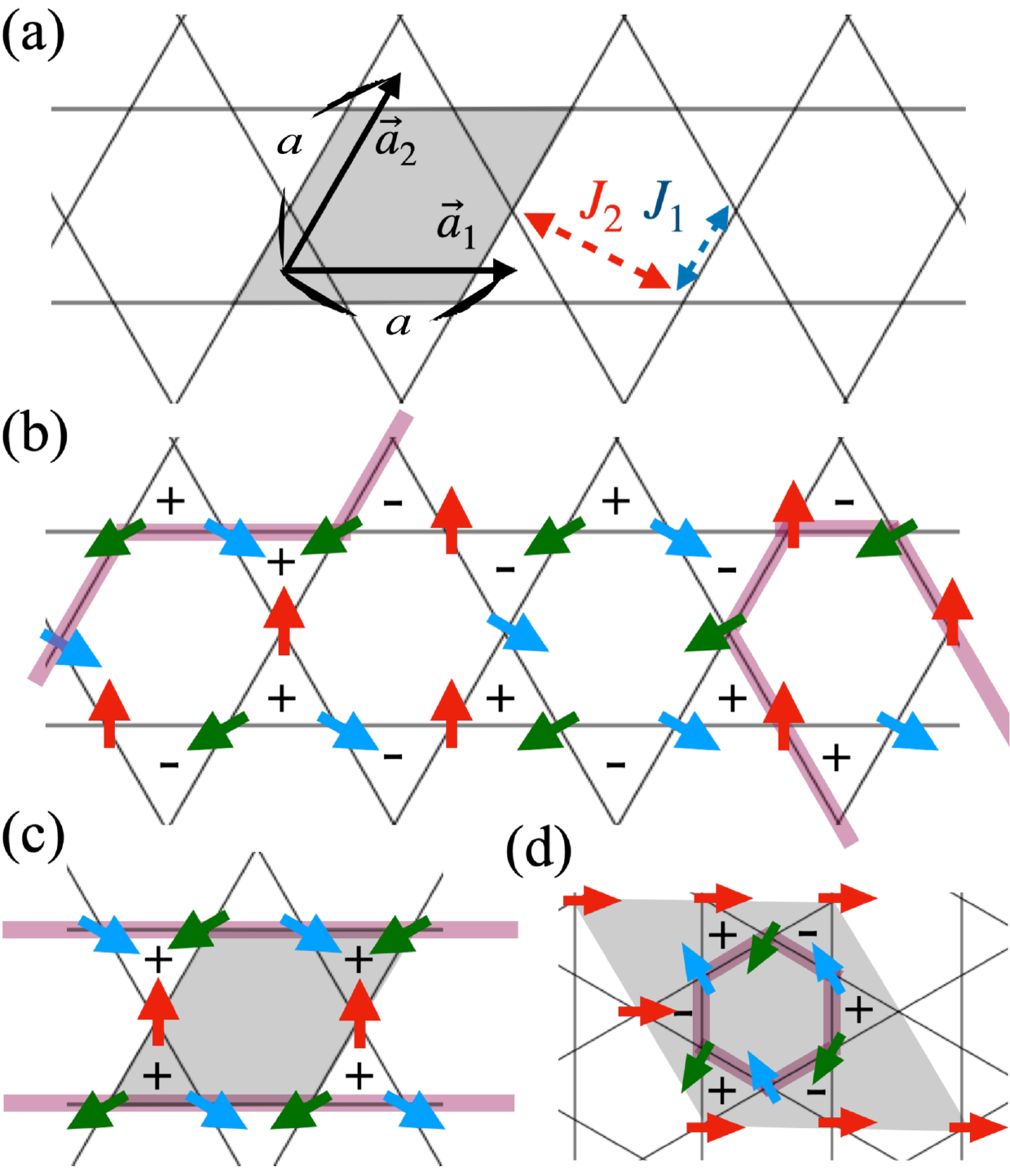}
   \caption{Schematic illustrations of the classical $J_1$-$J_2$ $XY$ kagome antiferromagnet and its ground states.
            (a) Primitive vectors, lattice constant $a~(=1)$, a unit cell, and nearest-neighbor ($J_1$) and next-nearest-neighbor ($J_2$) interactions. (b) Typical spin configuration of the octupole order at $J_2=0$. The $+$ and $-$ denote the signs of the z component of the vector chirality. The red lines denote typical closed ``loops'' (see Sec.~\ref{sec:orders}) under periodic boundary conditions. (c) Ferrochiral and $q=0$ magnetic order for $J_2 < 0$. (d) Antiferrochiral and \sqrtthreesqrtthree magnetic order for $J_2 > 0$.
            }
   \label{fig:model}
\end{figure}

A prototype of such frustrated spin models is the classical $J_1$-$J_2$ $XY$ antiferromagnet on the kagome lattice. 
Its Hamiltonian is defined as follows:
\begin{align}
   H = J_1\sum_{\ev{ij}}\vec{S}_i\cdot\vec{S}_j - J_2\sum_{\ev{\ev{ij}}}\vec{S}_i\cdot\vec{S}_j .
   \label{eq:H_j1j2}
\end{align}
Here, $\vec{S}_i=(S_{i}^{x}, S_{i}^{y})$ represents a unit vector at the site $i$.
$J_1~(=1)$ denotes the nearest-neighbor interactions, and $J_2$ denotes the next-nearest-neighbor ones [see Fig.~\ref{fig:model}(a)].
$\ev{ij}$ represents a pair of nearest neighboring sites, and $\ev{\ev{ij}}$ represents a pair of next-nearest neighboring ones. 
We note that the spontaneous symmetry breaking of continuous degrees of freedom in two-dimensional models with short-range interactions, such as the conventional magnetic long-range ordering (LRO) at finite temperatures, is prohibited by the Mermin-Wagner theorem~\cite{Mermin1966}.

We first review the basic properties of the $J_1$-$J_2$ $XY$ antiferromagnets.
When $J_2=0$ and $T=0$, although the local $120^\circ$ spin order occurs at each triangular, there is no constraint on the global covering of the local $120^\circ$ spin order as shown in Figure~\ref{fig:model}(b). 
Thus, magnetic orders are prohibited even at zero temperature because of the macroscopic degeneracy.
However, it is proposed that the higher-order multipole degrees of freedom, i.e., octupole degrees of freedom, can have the LRO~\cite{baxter1970colorings}.

When $J_2=0$ and $T > 0$, it is expected that the system undergoes a Berezinskii-Kosterlitz-Thouless (BKT) transition~\cite{berezinskii1971destruction, Kosterlitz1973} from an octupole quasi-long-range ordered (QLRO) phase to a paramagnetic phase~\cite{Huse1992}.
Previous Monte Carlo (MC) simulations have estimated the BKT transition temperature \TBKT to be $\TBKT = 0.070$--$0.076$~\cite{Rzchowski1997}.
Additionally, a recent tensor network calculation has also estimated $\TBKT \simeq 0.0755$~\cite{song2023tensor}, which is consistent with the results obtained from the MC simulations.

Compared with the case of $J_{2}=0$, less is known about the effects of finite $J_2$.
At $T = 0$, $J_2~(\neq 0)$ lifts the macroscopic ground-state degeneracy.
As a result, the $q=0$ state becomes the ground state for $J_2<0$, while the \sqrtthreesqrtthree state becomes the ground state for $J_2>0$~\cite{zeng1990numerical}. 
In the kagome lattice, the LRO of the $z$ component of the vector chirality $\kappa^z_i$ can be accompanied by the magnetic order, which is defined as
\begin{equation}
\kappa^z_i \equiv \frac{2}{3\sqrt{3}}(\vec{S}_{i_1}\times\vec{S}_{i_2} + \vec{S}_{i_2}\times\vec{S}_{i_3} + \vec{S}_{i_3}\times\vec{S}_{i_1})^z,   
\end{equation}
where $i_{n}$ represents the site index in the triangular [refer to Figs.~\ref{fig:model}(c) and (d), and Sec.~\ref{sec:orders}].
As shown in Figs.~\ref{fig:model}(c) and (d), ferrochiral (antiferrochiral) ordering is accompanied by the $q=0$ (\sqrtthreesqrtthree) magnetic order.

For $T>0$, the magnetic orders become QLROs, while the Ising-type chiral orders are anticipated to remain LROs.
Thus, the system undergoes the magnetic BKT transitions and the chiral long-range transitions to a paramagnetic phase at finite temperatures.
The magnetic BKT transition temperatures approach zero as $|J_2|$ decreases.
Additionally, the chiral transition temperatures also exhibit $J_2$ dependencies similar to the magnetic BKT transition temperatures.
In the case of the triangular lattice, there is a slight differentiation between the transition temperatures for the chiral transition and the antiferromagnetic BKT transition~\cite{Misawa2010, miyashita1984nature, Ozeki2003}.
However, this particular aspect remains unexplored for the kagome lattice.
Intriguing questions also arise regarding how these chiral orders dissolve at finite temperatures and the nature of the relationship between the chiral transitions and the BKT transitions of magnetic orders.

For the kagome lattice, both previous phenomenological~\cite{Korshunov2002} and numerical studies~\cite{song2023tensor} have proposed schematic phase diagrams for $J_2\neq 0$.
Additionally, for $J_2<0$, these studies proposed the emergence of a non-trivial first-order transition.  
Interestingly, a similar first-order transition was reported in MC simulations for the classical $J_1$-$J_2$ Heisenberg antiferromagnet~\cite{spenke2012classical}.
However, due to the numerical challenges arising from the ground-state degeneracy and low-temperature phase transitions, the outcomes of classical MC simulations and computed phase diagrams for the classical $J_1$-$J_2$ $XY$ kagome antiferromagnet have not been reported yet.

\begin{figure}[htb]
   \centering
   \includegraphics[width=1.0\columnwidth]{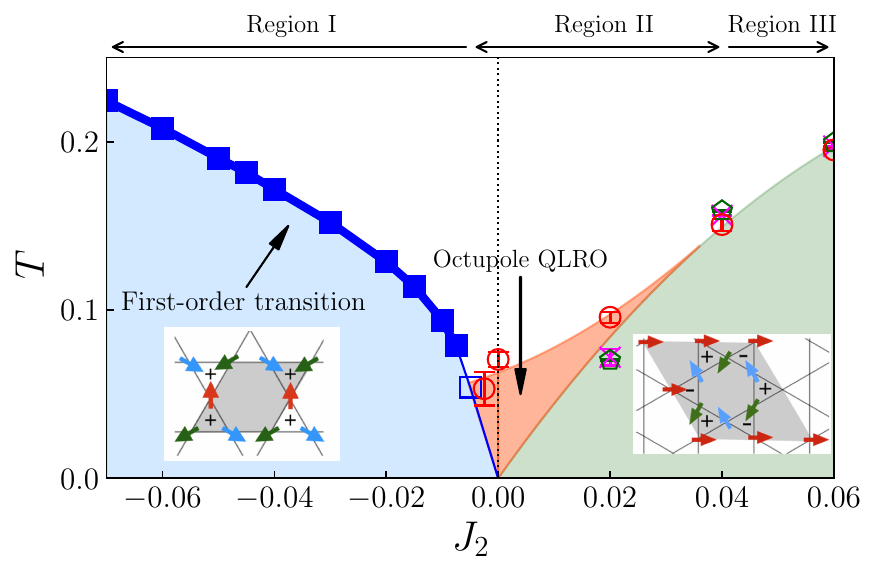}
   \caption{Computed $J_2$-$T$ phase diagram.
            The insets denote the ground-state spin configurations for $J_2 \neq 0$.
            The left one represents the ferrochiral and $q=0$ magnetic order [Fig~\ref{fig:model}(c)], while the right one represents the antiferrochiral and $\sqrtthree\times \sqrtthree$ magnetic order [Fig~\ref{fig:model}(d)].
            The thick blue line represents a first-order phase transition.
            There remain unresolved issues near the terminal point on the right side of the first-order transition line. 
            For details, please refer to Fig.~\ref{fig:proposed_phase_diagram} and discussions in the main text.}
   \label{fig:phase_diagram}
\end{figure}

The objective of this study is to quantitatively elucidate the cooperative effect of $T$ and $J_2$ on the macroscopic degeneracy of this model's ground state.
To this end, we map out a $J_2$-$T$ phase diagram (see Fig.~\ref{fig:phase_diagram}) by large-scale classical MC simulations applying both equilibrium MC and non-equilibrium relaxation (NER) methods~\cite{Ozeki2003}. 
We also reveal the existence of a non-trivial first-order transition when $J_2<0$.

This paper is structured as follows: 
Section~\ref{sec:orders} provides a more detailed explanation of the magnetic and chiral orders. 
Section~\ref{sec:method} describes the MC methods used in our study. 
In Sec.~\ref{sec:results}, we present the computed phase diagram, as well as the MC data for the phase diagram. 
Section \ref{sec:nature} is dedicated to the discussion of the nature of the first-order transition. 
Finally, in Sec.~\ref{sec:summary}, we summarize the main results of this study.

\section{Magnetic and chiral orders}\label{sec:orders}
In this section, we summarize definitions of the lattice structure, magnetic, and chiral orders.

\subsection{Unit cell}
As illustrated in Fig.~\ref{fig:model}(a), we use the primitive vectors given by
\begin{align}
   \vec{a}_1 &= (a, 0), \\
   \vec{a}_2 &= (a/2, \sqrt{3}a/2),
\end{align}
where $a~(=1)$ represents the lattice constant.
Each unit cell contains one upward-facing triangle, which consists of three sites [see the unit cell in Fig.~\ref{fig:model}(a)].
In the following numerical simulations, we consider periodic systems of square geometry with $L^2$ unit cells, including $3L^2$ spins.

\subsection{Local $120^\circ$ structure}\label{sec:local120}
In this subsection, we explain the local $120^\circ$ structure, 
a building block for the magnetic orders.
As illustrated in Fig.~\ref{fig:model}(b), three spins on each triangle are apart from each other by $120^\circ$.
The three colors indicate the three spins pointing in different directions.
This configuration is called the local $120^\circ$ structure. 
Only this local $120^\circ$ structure satisfies 
the sum rule $\vec{S}_1+\vec{S}_2+\vec{S}_3=0$ up to a global rotation.

All states with local $120^\circ$ structures on every triangle minimize the Hamiltonian for $J_2 = 0$.
This can be seen by rewriting the Hamiltonian as follows:
\begin{align}
   H = \sum_{\mathrm{triangle}} (\vec{S}_1+\vec{S}_2+\vec{S}_3)^2 + \mathrm{const.},
   \label{eq:H_j1}
\end{align}
where the sum runs over all upward-facing and downward-facing triangles.
The number of such ground states grows exponentially with the number of sites, resulting in magnetic disorder~\cite{baxter1970colorings, Huse1992}.

As illustrated in Fig.~\ref{fig:model}(b), a local $120^\circ$ state can have either $\kappa^z=1$ or $-1$.
Therefore, the (anti)ferrochiral order is expected to be realized when finite $J_{2}$ induces a periodic order of the local $120^\circ$ structures, such as the $q=0$ magnetic order and the $\sqrt{3}\times\sqrt{3}$ magnetic order. 

In the $120^\circ$ state, a ``loop'' can be formed by spins alternating two out of three colors, as shown in Fig.~\ref{fig:model}(b).
The two types of spins on such a loop can be interchanged without any energy loss since this does not violate the sum rule. 
From now on, we define the length of a loop as the number of sites it contains.
At finite $T$, a loop becomes not well-defined since spins deviate from perfect local $120^\circ$ structures due to thermal fluctuations. 
In the present study, for the loop update described in Sec.~\ref{sec:method} and the discussion about the nature of the first-order transition in Sec.~\ref{sec:nature}, we define a loop at finite $T$ using the algorithm proposed in Ref.~\onlinecite{Schnabel2012}.

\subsection{Octupole order ($J_2=0$)}
In this subsection, we explain the octupole order at $J_2=0$ and $T=0$.
As mentioned above, at $J_{2}=0$, all spin states consisting of the local $120^\circ$ structures have the same energy [a typical spin state is shown in Fig~\ref{fig:model}(b)].
Consequently, these states have neither magnetic nor chiral LRO.
However, composite degrees of spins, termed octupole, can exhibit LRO at $T=0$.

Additionally, in the context of loops, an octupole order does not exhibit any periodicity similar to those observed for the $q=0$ [Fig~\ref{fig:model}(c)] and the \sqrtthreesqrtthree order [Fig~\ref{fig:model}(d)].

The octupole order parameter is defined as follows~\cite{Rzchowski1997}:
\begin{align}
   m^2_{\mathrm{oct}} \equiv 
   \frac{1}{N^2}\left[ \left(\sum_i \cos 3\theta_i \right)^2+\left(\sum_i\sin 3\theta_i\right)^2\right],
   \label{eq:m2_octupole}
\end{align}
where $\theta_i$ represents the angle of $i$th spin measured from the $x$ axis and $N$ is the number of spins.

\subsection{$q=0$ and ferrochiral order ($J_2<0$)}
In this subsection, we explain the $q=0$ and ferrochiral orders for $J_2<0$.
This state is illustrated in Fig.~\ref{fig:model}(c).
The spin configuration is translationally invariant; thus, $\kappa^z$ has the same sign on all the triangles.
This indicates that the $q=0$ magnetic order coexists with the ferrochiral order at $T=0$.
At $T \neq 0$, the ferrochiral LRO can survive because it is a spontaneous symmetry breaking of discrete degrees of freedom, while the $q=0$ magnetic order becomes QLRO.

As shown in Fig.~\ref{fig:model}(c), the spin configuration has a one-dimensional ``loop'', whose length is of the order of the system size $L$.
Such a loop is longer than that of the octupole-ordered states.

The magnetic and chiral order parameters are defined as follows:
\begin{align}
   m^2_{q=0} &= 
   \frac{1}{3N^2_{\Delta}}\sum_{l=1}^3\left(\sum_{i=1}^{N_{\Delta}} \vec S_l^i \right)^2,\label{eq:m2_q0}\\
   \kappa_{\mathrm{Ferro}}^2  &= 
   \left( \frac{1}{N_{\Delta}}\sum_{\mathrm{all}\Delta}\kappa_{\Delta} + \frac{1}{N_{\nabla}}\sum_{\mathrm{all}\nabla}\kappa_{\nabla}\right)^2,
   \label{eq:chiral_fm}\\
   \kappa_{\Delta,\nabla} &= \frac{2}{3\sqrt{3}}(\vec S_1 \times \vec S_2 + \vec S_2 \times \vec S_3 + \vec S_3 \times \vec S_1)_{\Delta,\nabla},
\end{align}
where the index $i$ represents an $i$th upward-facing triangle and $l$ denotes the $l$th site on each triangle.
Furthermore, $N_{\Delta}$ and $N_{\nabla}$ denote the number of upward- and downward-facing triangles, respectively.

\subsection{\sqrtthreesqrtthree \ and antiferrochiral order ($J_2>0$)}
Figure~\ref{fig:model}(d) illustrates the \sqrtthreesqrtthree order at $T=0$ for $J_2 > 0$, whose unit cell is larger than that of the $q=0$ order.
In this spin configuration, $\kappa^z$ has different signs in upward- and downward-facing triangles.
The \sqrtthreesqrtthree order coexists with the antiferrochiral order at $T=0$.

As shown in Fig.~\ref{fig:model}(d), each unit cell contains a loop consisting of spins with alternating two colors (green and blue in the figure).
The length of the loop is 6, which is the shortest possible length of a closed loop.

The magnetic and chiral order parameters are defined as follows:

\begin{eqnarray}
   m^2_{\mathrm{\sqrt3\times\sqrt3}} &=& 
   \frac{1}{3N^2_{\Delta}}\sum_{l=1}^{3}\left[\sum_{i=1}^{N_{\Delta}} \vec S_l^i \exp\left( \frac{2\pi\mathrm{i}}{3}(x_l^i+y_l^i)\right)\right]^2,\label{eq:m2_sqrt3}\\
   \kappa_{\mathrm{AF}}^2 &=& 
   \left(\frac{1}{N_{\Delta}}\sum_{\mathrm{all}\Delta}\kappa_{\Delta} - \frac{1}{N_{\nabla}}\sum_{\mathrm{all}\nabla}\kappa_{\nabla}\right)^2.
   \label{eq:chiral_afm}
\end{eqnarray}
Here, $x^i_l$ and $y^i_l$ is defined as $\vec{r}^i_l = x^i_l\vec{a}_1 + y^i_l\vec{a}_2$ where $\vec{r}^i_l$ denotes real-space position of the $l$th site in the $i$th upward-facing triangle.

\section{Method}\label{sec:method}
In this section, we explain the classical MC methods used in this study for numerical calculations.
To reveal the thermodynamic properties of the model, we employed two complementary methods: The standard equilibrium MC method (for studying the long-time limit of small systems) and the NER method (for studying the short-time relaxation process of large systems).
In the following subsections, we explain these methods in detail.

\subsection{MC simulations of equilibrium systems}
To simulate equilibrium states and avoid the freezing of MC dynamics, we used several update methods:
the random-flip update~\cite{marsaglia1972choosing}, the over-relaxation update~\cite{Creutz1987, Alonso1996}, the gaussian-move update~\cite{hinzke1999monte, Evans2014} and the non-local loop update~\cite{Schnabel2012}.
We further employed the replica exchange Monte Carlo method for efficient multiple-temperature simulations~\cite{hukushima1996exchange}.

In particular, the non-local loop update~\cite{Schnabel2012} takes advantage of the fact that, at $T=0$ and $J_2=0$, one can interchange the types of spins on a loop consisting of two alternating colors without any energy loss.
This enables us to simultaneously update spins on a loop, resulting in transitions between degenerate ground states, even at low temperatures where other local updates are frozen.
However, at finite $T$, a loop is not well-defined since spins deviate from perfect local $120^\circ$ structures due to thermal fluctuations.
In actual calculations, we construct a loop as follows:
At first, we randomly select two nearest-neighboring spins denoted as $\vec{S}_{l_1}$ and $\vec{S}_{l_2}$, which represent the first and second spins on a loop.
Next, we calculate the inner product between $\vec{S}_{l_1}$ and each of nearest-neighboring spins of $\vec{S}_{l_2}$ excluding $\vec{S}_{l_1}$.
The spin with the largest inner product is chosen as $\vec{S}_{l_3}$ so that $\vec{S}_{l_1}$ and $\vec{S}_{l_3}$ oriented in approximately the same direction.
This procedure is repeated until the loop is closed (see Ref.~\onlinecite{Schnabel2012} for more details).
Each attempt of loop update is accepted with the probability depending on the total energy change by the standard Metropolis algorithm.

One MC step involves one sweep through the system with local updates and the non-local loop update, followed by attempts of replica exchanges between neighboring temperatures. 
After thermalization, we evaluate the physical quantities defined in Sec.~\ref{sec:orders}. 
We typically take $10^6$ MC steps for the thermalization and  $9\times10^6$ MC measurement steps.
The physical quantities are measured every 10 MC steps.

\subsection{NER method}
In the NER analyses, we study the relaxation processes of a large system from an initially ordered state using only the single-spin updates (both the loop update and the replica exchange are disabled).
In this study, time $t$ is measured in units of MC steps.

First, as an initial state ($t=0$), we choose a specific ordered state whose transition temperature we want to estimate.
In the relaxation process, we evaluate the dynamical correlation function defined as follows:
\begin{align}
   G_{O}(t) \equiv \ev{O(0) \cdot O(t)}
   \label{eq:dym_corr_func}
\end{align}
for the order parameter $O$ of interest (i.e., corresponding to the ordering).
Here, $O(t)$ is the value of the order parameter at $t$.
The symbol $\ev{\cdots}$ denotes the sample average over MC results with different random number seeds, which is taken to suppress statistical fluctuations due to finite system sizes.

For $T > T_{\mathrm{BKT}}$, $G(t)$ is expected to decay exponentially as follows:
\begin{align}
   G(t) =  a \ \mathrm e^{-t/\tau}.
   \label{eq:time_corr_func}
\end{align}
Here, $\tau$ denotes a temperature-dependent relaxation time.
Instead of estimating $\tau$ at each $T$ using the least-squares method with this equation, we employ the scaling analysis~\cite{Ozeki2003} detailed in Appendix~\ref{sec:scaling} because $G(t)$ does not decay exponentially at short times.
The scaling analysis allows us to determine the $T$ dependence of $\tau$ simultaneously.
Once the $T$ dependence of $\tau$ is determined, we estimate the transition temperature by assuming~\cite{Kosterlitz1973, Kosterlitz1974}
\begin{align}
   \tau = b \ \exp [\frac{c}{\sqrt{T-T_{\mathrm{BKT}}}}].
\end{align}
By taking the logarithm of this equation, we get
\begin{align}
   \ln{\tau} = \ln{b} + \frac{c}{\sqrt{T-T_{\mathrm{BKT}}}},
   \label{eq:T_BKT_from_logtau}
\end{align}
which can be used to estimate $T_{\mathrm{BKT}}$ by the least-squares method.

Since the chiral degrees of freedom can show the long-range order, the temperature dependence of $\tau$ around the chiral transition is anticipated to show the following power-law behavior, which is defined as
\begin{align}
   \tau = b \ (T-T_{\mathrm{c}})^{-z\nu},
\end{align}
where $z$ is the dynamical critical exponent and $\nu$ is the critical exponent of the correlation length~\cite{nishimori2010elements}.

Using the standard Metropolis single-spin update, we typically perform the relaxation up to $10^5$ MC steps for the system size $3L^2$ up to $L=1800$ (one sample/MC run). 
This process takes over 24 hours with one core of AMD EPYC 7702P.
The relaxation time $\tau$ is determined at up to 11 different temperatures.
The sample averages are computed from as many as 300 independent MC runs for each temperature.
For our initial state, we use $q=0$ state for $J_2<0$ and the \sqrtthreesqrtthree state for $J_2>0$, both of which are perfectly ordered.
Additionally, we use \sqrtthreesqrtthree initial state for $J_2=0$ because the classical Heisenberg antiferromagnet on the kagome lattice is expected to have the \sqrtthreesqrtthree antiferromagnetic LRO in the $T\rightarrow 0$ limit~\cite{Huse1992}.
We confirmed that the finite-size effect is negligibly small in all the calculations shown in this paper.

\section{Results}\label{sec:results}
We start our discussion by providing an overview of the computed $J_2$-$T$ phase diagram in Sec.~\ref{sec:pd}.
In the three subsequent subsections, we discuss MC results for three different regions of $J_2$.

\begin{figure*}
   \begin{center}
      \includegraphics[width=0.99\linewidth]{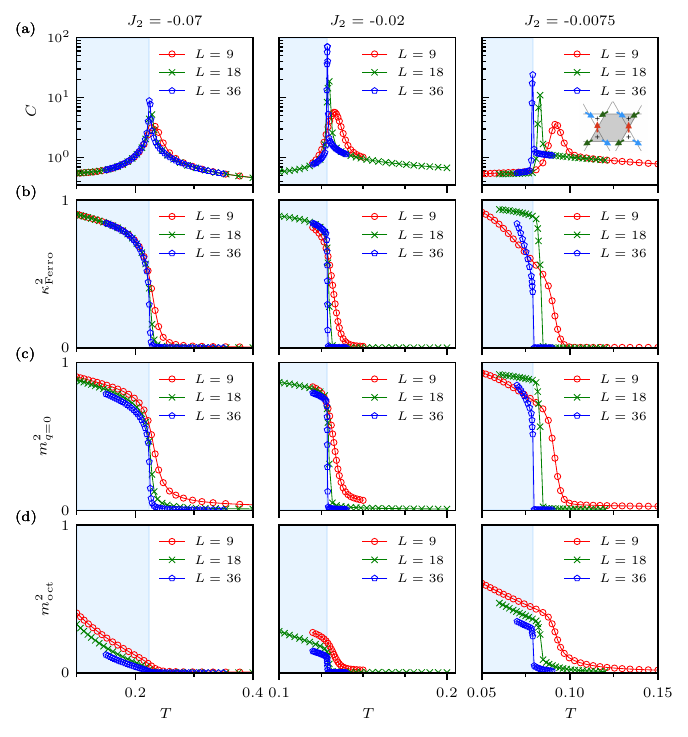}
   \end{center}
   \caption{Temperature dependence of the specific heat [(a)] and the order parameters \fchiral [(b)], \mqzero [(c)], \moctupole [(d)] computed at $J_2=-0.07, -0.02, -0.0075$ in Region I. 
            The system sizes were $L=9$, $18$, and $36$. 
            The vertical lines denote the first-order transition temperature at the bulk limit estimated by the finite-size scaling analyses~\cite{challa1986finite} (see Fig.~\ref{fig:Tc_scaling_1storder}).
   }
   \label{fig:eqvals_region1}
\end{figure*}

\begin{figure}
   \centering
   \includegraphics[width=0.75\columnwidth]{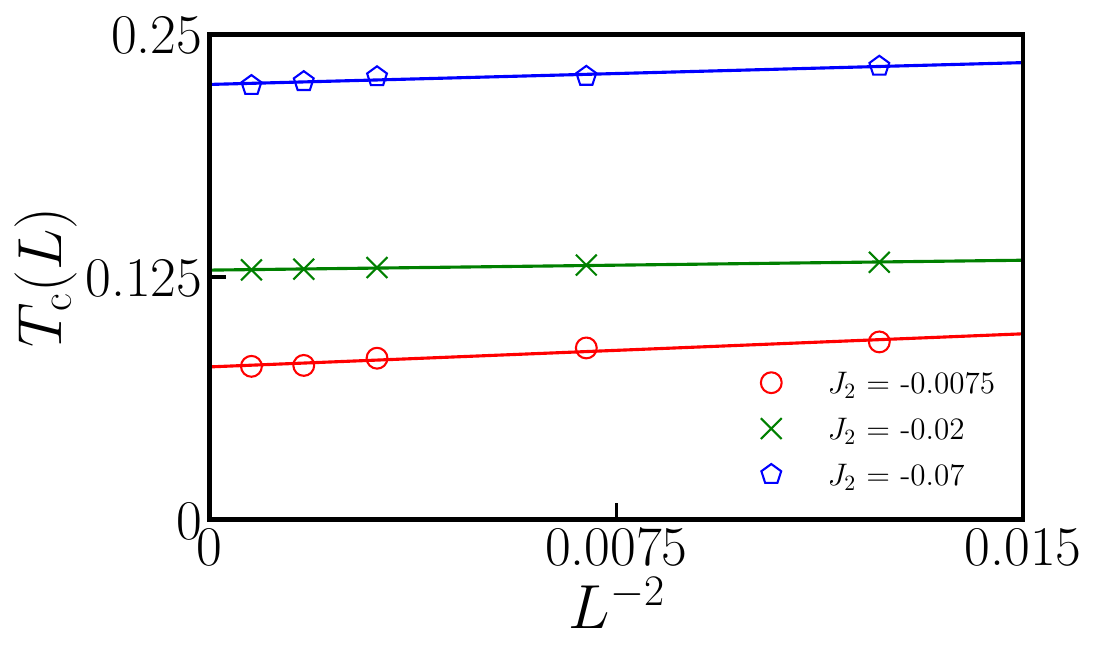}
   \caption{The size dependence of $T_{\mathrm{c}}(L)$ defined as the temperature exhibiting the specific-heat peak for each $L$.
           They are computed at $J_2=-0.0075, -0.02$, and -0.07.}
   \label{fig:Tc_scaling_1storder}
\end{figure}

\begin{figure}
   \centering
   \includegraphics[width=0.7\columnwidth]{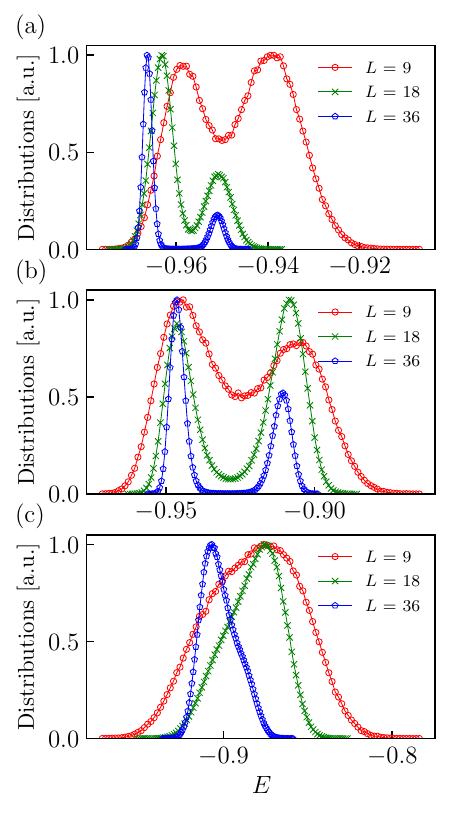}
   \caption{Energy histogram measured at $J_2 = -0.0075$ [(a)], $-0.02$ [(b)] and $-0.07$ [(c)].}
   \label{fig:energy_histogram}
\end{figure}

\subsection{Overview of the $J_2$-$T$ phase diagram}~\label{sec:pd}
Figure~\ref{fig:phase_diagram} shows the $J_2$-$T$ phase diagram.
The overall structure at low temperatures is consistent with previous studies~\cite{Korshunov2002, song2023tensor}.
Specifically, for $J_2 < 0$, a phase exists where the ferrochiral LRO and the $q=0$ magnetic QLRO coexist (the blue region in the figure).
For $J_2>0$, a phase exists where the antiferrochiral LRO and the $\sqrt{3}\times\sqrt{3}$ magnetic QLRO coexist (the green region in the figure).
Between these two phases, there exists a phase with the octupole QLRO (the red region in the figure).

Next, we discuss temperature-driven phase transitions.
Near $J_2=0$, there is a BKT transition between the octupole QLRO and the paramagnetic phases.
For $J_2>0$, an antiferrochiral transition and a \sqrtthreesqrtthree BKT transition seem to occur nearly at the same temperature. 
These two transition temperatures are indeed slightly separated, as we will discuss in greater detail.

For $J_2<0$, there are even richer structures.
In particular, there is a first-order transition between the ferrochiral LRO and paramagnetic phases, which is a remarkable finding in this study.
The first-order transition line seems to terminate near the intersection where the octupole BKT transition line reaches the ferrochiral LRO phase.

In the following subsections, we show numerical data for the three distinct regions of $J_2$:
Region I ($J_2 \le -7.5\times10^{-3}$) involving the first-order transition,
Region II with the octupole phase ($-7.5\times10^{-3}<J_2< 4\times10^{-2}$),
Region III with the coexisting ferrochiral LRO and \sqrtthreesqrtthree QLRO ($J_2>4\times10^{-2}$).

\subsection{Region I~($J_2/J_1\le -7.5\times10^{-3}$)}\label{sec:regionI}
In this subsection, we delve into the equilibrium MC results for Region I, where the first-order transition exists between the ferrochiral LRO and paramagnetic phases.
According to the ground-state phase diagram for $J_2<0$, the following three order parameters are expected to be relevant in this region: \mqzero [Eq.~\eqref{eq:m2_q0}], \fchiral [Eq.~\eqref{eq:chiral_fm}], and \moctupole [Eq.~\eqref{eq:m2_octupole}].
The Mermin-Wagner theorem indicates that the order parameters \mqzero and \moctupole vanish at the bulk limit $L \rightarrow \infty$.
Nevertheless, the size dependence of  \mqzero and \moctupole can give us essential insights into the nature of the QLRO.

Figure~\ref{fig:eqvals_region1} shows the temperature dependence of the specific heat and the order parameters computed for $L=9$, $18$, and $36$.
Let us first discuss the specific heat as shown in Fig.~\ref{fig:eqvals_region1}(a).
At all three values of $J_2=-0.0075$, $-0.02$, and $-0.07$, the temperature dependence of the specific heat displays a singular peak, which becomes sharper as $L$ increases.
This result indicates the existence of a single first-order phase transition from the paramagnetic phase to the low-$T$ phase with the coexisting ferrochiral LRO and $q=0$ magnetic QLRO.
The transition temperature of the first-order phase transition increases when $J_2$ changes from $-0.0075$ to $-0.07$.
We note that the intermediate ferrochiral LRO phase without the magnetic QLRO may appear in the case of $J_{2}<0$, but it is difficult to discuss the existence of the intermediate phase within the accuracy of the current calculations.

To further examine the existence of the first-order phase transition, we performed the finite-size scaling analyses of $T_{\mathrm{c}}(L)$ at which the specific heat exhibits its maximum.
For a first-order transition, $T_{\mathrm{c}}(L)$ is anticipated to be proportional to $L^{-2}$~\cite{challa1986finite}.
As shown in Fig.~\ref{fig:Tc_scaling_1storder}, for all $J_2$ values, the scaling of $T_{\mathrm{c}}(L)$ seems to agree well with anticipated $L^{-2}$ behavior.
This result suggests the existence of the first-order phase transition.

We also examine the energy histogram at $T_{\mathrm{c}}$.
As shown in Figs.~\ref{fig:energy_histogram}(a) and (b), the energy histogram shows the double-peak structure and it becomes more pronounced as $L$ increases. 
These observations indicate the first-order nature of the transition at $J_2=-0.0075$ and $-0.02$.
In contrast, at $J_2=-0.07$ [Fig.~\ref{fig:energy_histogram}(c)], the histogram exhibits only a single broad peak with a small shoulder structure even for the largest system ($L=36$).
This indicates that the transition may be continuous.

In order to better understand the nature of the low-$T$ phase, we analyze the MC data for the order parameters.
As shown in Fig.~\ref{fig:eqvals_region1}(b), it is evident that the low-$T$ phase has ferrochiral LRO.
The jump-like behavior in \fchiral around $T_{\mathrm{c}}$ at $J_2=-0.02$ further supports the first-order nature of the transition.
Figure~\ref{fig:eqvals_region1}(c) shows the temperature dependence of \mqzero, which appears similar to that of \fchiral.
Consistent with analyses of the specific heat and the energy histograms, the jumps in the physical quantities become small by changing from $J_{2}=-0.02$ to $J_{2}=-0.07$. 
The steep changes in \mqzero and \fchiral around $T_{\mathrm{c}}$ at $J_{2}=-0.0075$ are consistent with the weak first-order phase transition.

Finally, we discuss the temperature dependence of \moctupole shown in Fig.~\ref{fig:eqvals_region1}(d). 
As previously mentioned, \moctupole becomes finite in the presence of the ferrochiral LRO.
As $T$ increases, both \moctupole and \fchiral seem to disappear at the same temperature within the accuracy of the current simulations.
This observation indicates that the non-zero value of \moctupole originates from the primary $q=0$ QLRO. 

\subsection{Region II ($-7.5\times10^{-3}<J_2/J_1<4\times10^{-2}$)}\label{sec:regionII}
In this subsection, we discuss the equilibrium MC results 
and the NER results for Region II.

\begin{figure*}
   \begin{center}
      \includegraphics[width=0.99\linewidth]{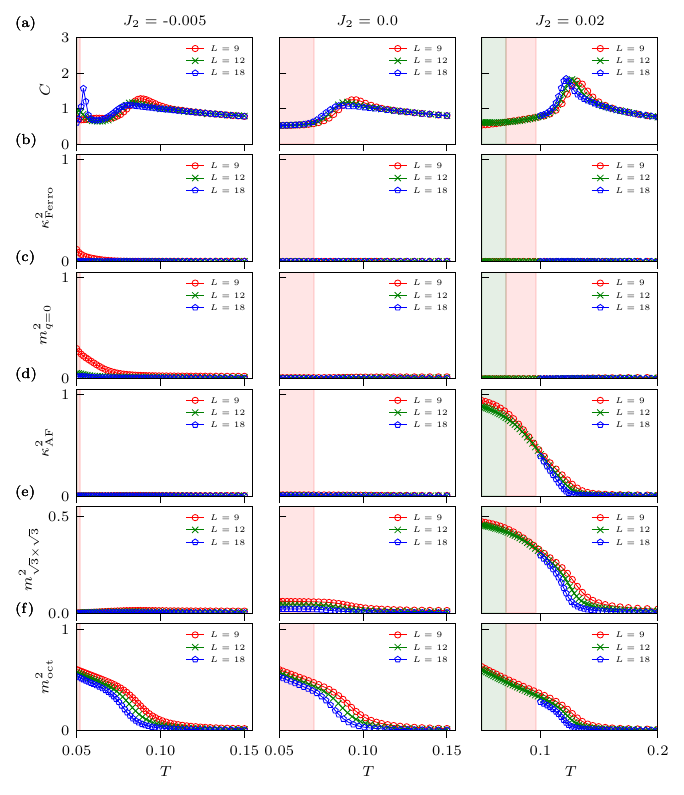}
   \end{center}
   \caption{
            Temperature dependence of the specific heat [(a)] and the order parameters \fchiral [(b)], \mqzero[(c)], \afchiral [(d)], \mqsqrtthree [(e)], \moctupole [(f)] computed at $J_2 = -0.005, 0, 0.02$ in Region II. 
            The system sizes were $L=9$, 12, and 18.
      }
   \label{fig:eqvals_region2}
\end{figure*}

\subsubsection{Results of equilibrium MC}
Figure~\ref{fig:eqvals_region2} shows the specific heat and various order parameters computed at $J_2=-0.005$, 0, 0.02.
We first discuss the results for a small but negative $J_{2}$ ($J_{2}=-0.005$), where the ground state is expected to be in the $q=0$ ordered phase.
The left columns of Fig.~\ref{fig:eqvals_region2} illustrate the temperature dependence of several physical quantities for $J_2=-0.005$.
Figure~\ref{fig:eqvals_region2}(a) depicts the specific heat, showing two peaks: the broad high-$T$ peak and the sharp low-$T$ peak.
As $L$ increases, the high-$T$ peak decreases in height, while the low-$T$ peak becomes more pronounced and sharpens.
As shown in Fig.~\ref{fig:eqvals_region2}(f), the position of the high-$T$ peak is near the onset of $\moctupole$, indicating that the high-$T$ peak in the specific heat corresponds to the transition from the paramagnetic phase to the octupole QLRO phase.

\begin{figure}[htb]
   \centering
   \includegraphics[width=0.9\columnwidth]{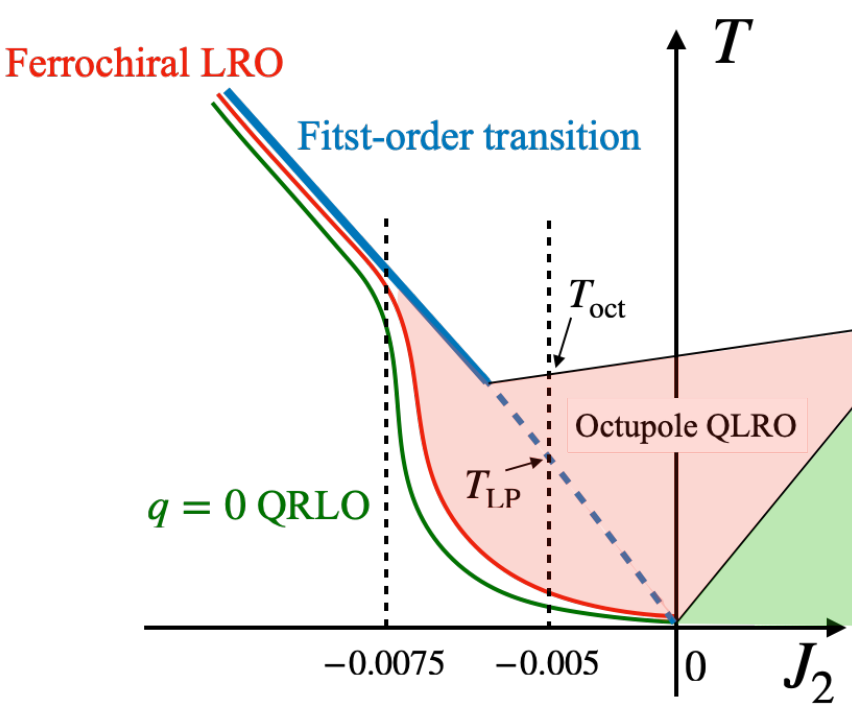}
   \caption{Proposed conceptual $J_2$-$T$ phase diagram near $J_2=0$. 
            The blue solid line indicates the first-order transition induced by a change in loop length from $O(L)$ to $O(1)$.
            The red solid line denotes the ferrochiral transition.
            The green solid line represents the $q=0$ magnetic BKT transition.
            $T_{\mathrm{LP}} (\simeq 0.06)$ represents the temperature at which the specific heat exhibits a low-$T$ peak for $L=18$, as shown in the left panel in Fig.~\ref{fig:eqvals_region1}(a).
   }
   \label{fig:proposed_phase_diagram}
\end{figure}

Below the low-$T$ peak, as shown in Figs.~\ref{fig:eqvals_region2}(b) and (c), $\fchiral$ and $\mqzero$ increase for $L=9$.
However, for $L=12$ and 18, $\fchiral$ and $\mqzero$ remain approximately zero below the low-$T$ peak down to $T=0.05$.
As illustrated in a conceptual phase diagram in Fig.~\ref{fig:proposed_phase_diagram}, this result indicates that the phase transitions of both the ferrochiral LRO and the $q=0$ magnetic QLRO may be separated from the first-order phase transition.
Based on the expectation, the ferrochiral LRO and the $q=0$ magnetic QLRO should appear for $T<0.05$.
To confirm the existence of these orders at low temperatures, we conducted equilibrium MC simulations and NER analyses. 
However, due to the system-size limitation in the equilibrium MC simulations and the excessively long relaxation time in NER processes, we were unable to identify these orders.
A more detailed analysis to examine the validity of the expected phase diagram will be a subject for future study.

We next discuss the results for $J_{2}=0$.
As shown in Fig.~\ref{fig:eqvals_region2}(a), the specific heat exhibits only a single broad peak.
This peak seems to coincide with the onset of \moctupole [Fig.~\ref{fig:eqvals_region2}(d)].
This result indicates that the octupole QLRO occurs at $J_{2}=0$.
As we will show later, using the NER, we estimate the octupole BKT transition temperature as $T_{\rm oct} = 0.071 \pm 0.005$.
This estimate is consistent with the previous results that there is only an octupole BKT transition at $T = 0.070$--$0.076$~\cite{lee1986, Rzchowski1997, Korshunov2002, song2023tensor}.
Additionally, as shown in Fig.~\ref{fig:eqvals_region2}(e), \mqsqrtthree has small but non-zero values below the BKT transition temperature.
This result is consistent with the proposal that the \sqrtthreesqrtthree pattern is selected in the $T\rightarrow 0$ limit for the classical Heisenberg antiferromagnet on the kagome lattice~\cite{Huse1992}.

Lastly, we discuss the results for $J_{2}=0.02$, where the ground state is expected to be the $\sqrtthreesqrtthree$ ordered phase.
As shown in Fig.~\ref{fig:eqvals_region2}(a), the specific heat exhibits a single broad peak similarly at $J_2=0$ indicating the octupole QLRO.
In contrast, the low-$T$ behavior of the order parameters [Figs.~\ref{fig:eqvals_region2}(d)--(f)] is distinctly different from $J_2=0$.
In particular, both $\afchiral$ and $\mqsqrtthree$ become enhanced at low $T$, suggesting the coexistence of the antiferrochiral LRO and $\sqrtthreesqrtthree$ magnetic QLRO. 
Since it is difficult to accurately estimate the transition temperatures of the octupole QLRO and the magnetic QLRO by the equilibrium MC method, we perform the NER analysis for larger system sizes.
As we will discuss later, the transition temperatures of the octupole QLRO and magnetic QLRO are close to each other but are separated.

\subsubsection{Results of NER}
\begin{figure*}[htb]
  \begin{center}
   \includegraphics[width=0.9\textwidth]{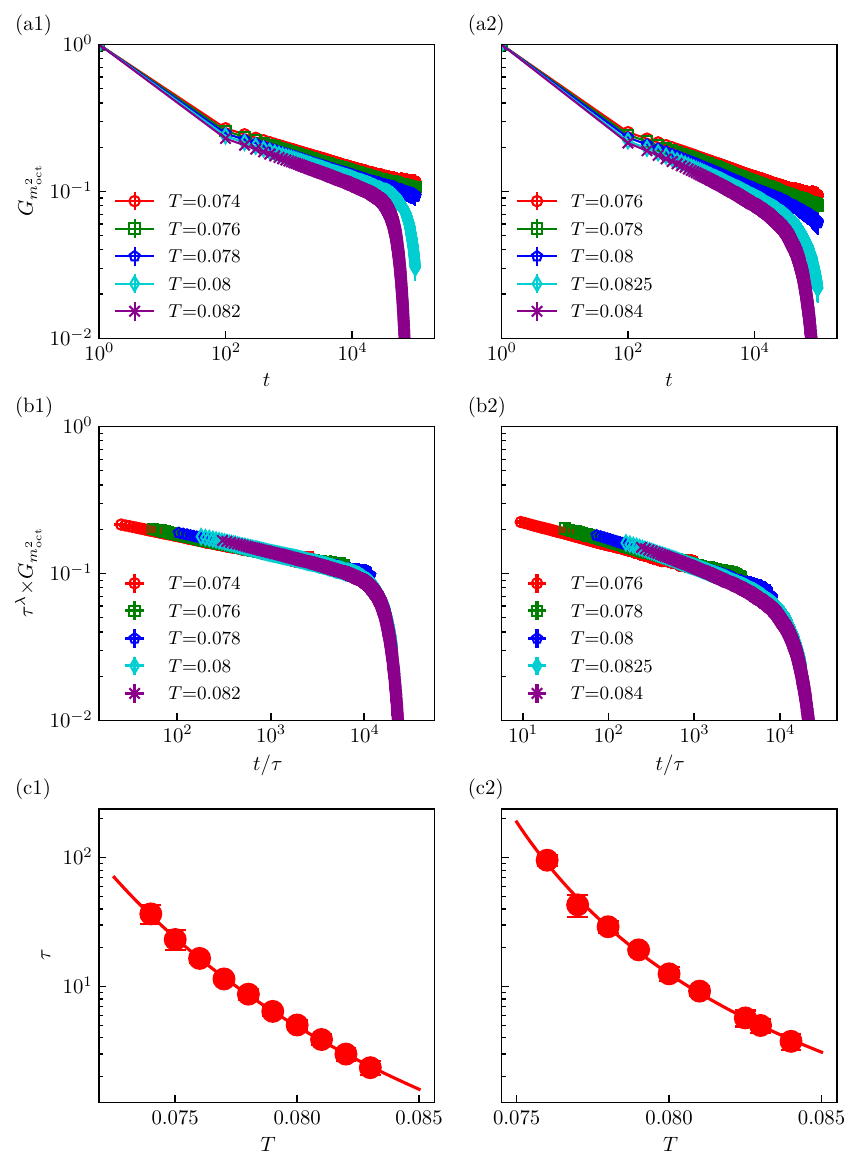}
   \caption{NER results for $J_2 = -0.0025$ (shown in the left panel) and $J_2 = 0$ (shown in the right panel). 
            The system sizes are $L = 1800$.
            The dynamical correlation functions $G(t)$ are computed solely for the octupole order parameter \moctupole. 
            (a) $G(t)$ as a function of the MC step $t$.
            (b) Scaling plot for $G(t)$.
            (c) The relaxation time $\tau$ as a function of temperature $T$ in an arbitrary unit. 
                The curve presents a fit by BKT scaling $\tau = b \ \exp (c/\sqrt{T-T_{\mathrm{BKT}}})$.
   }
  \label{fig:ner_j2_-2.5e-3_0}
  \end{center}
\end{figure*}

\begin{figure*}[htb]
  \begin{center}
   \includegraphics[width=0.9\textwidth]{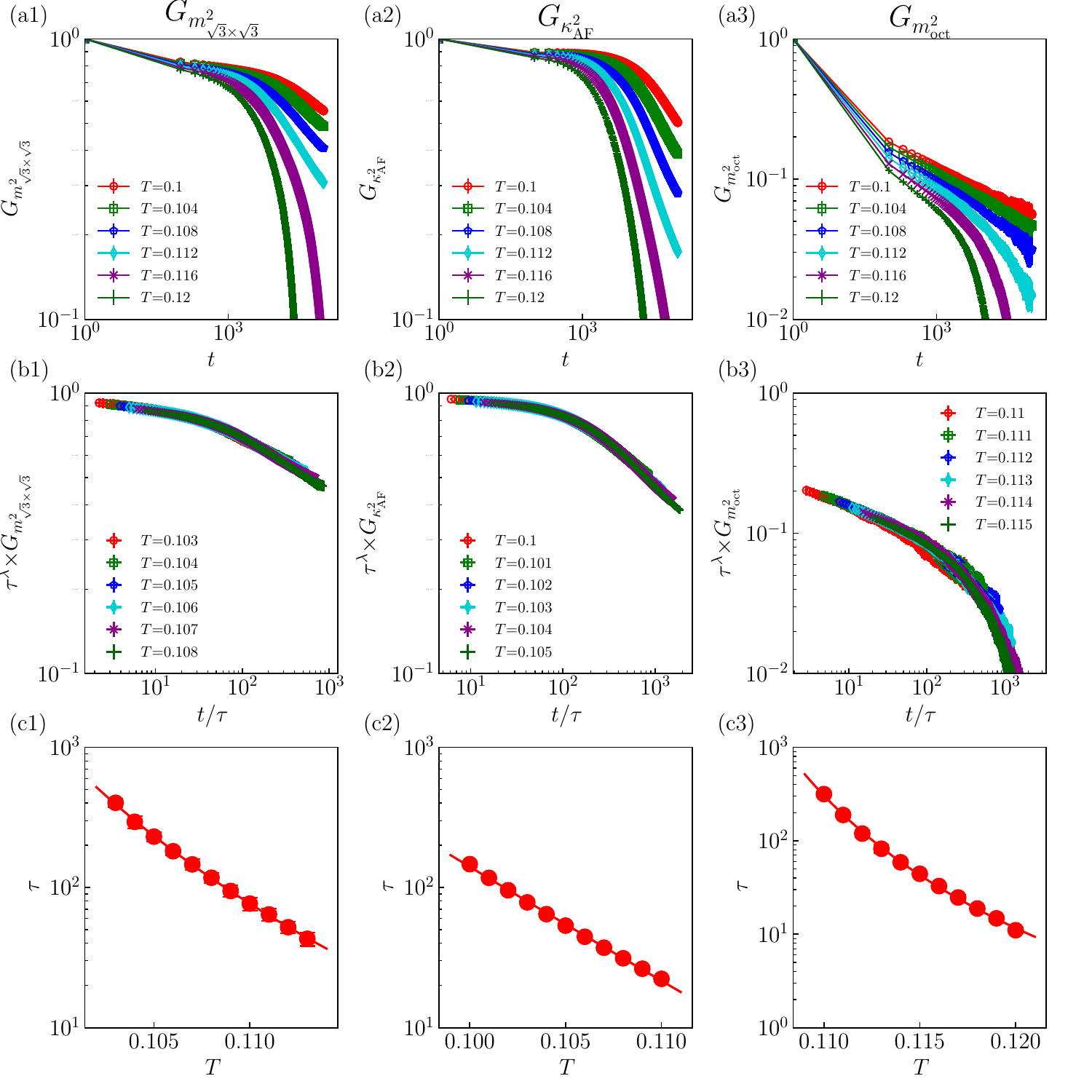}
   \caption{NER results for $J_2 = 0.02, L = 1800$. 
            The dynamical correlation function $G(t)$ is computed for the \sqrtthreesqrtthree order parameter \mqsqrtthree, the antiferrochiral order parameter \afchiral, and the octupole order parameter \moctupole.
            (a) $G(t)$ as a function of the MC step $t$.
            (b) Scaling plot for $G(t)$.
            (c) The relaxation time $\tau$ as a function of temperature $T$ in an arbitrary unit. 
                The curve presents a fit by BKT scaling $\tau = b \ \exp (c/\sqrt{T-T_{\mathrm{BKT}}})$ in (c1) and (c3) and by the power law $\tau = b (T-T_{\mathrm{c}})^{-z\nu}$ in (c2).
   }
  \label{fig:ner_j2_0.02}
  \end{center}
\end{figure*}

In this subsection, we present the results obtained by the NER method.
Figures~\ref{fig:ner_j2_-2.5e-3_0} and \ref{fig:ner_j2_0.02} illustrate the results of $J_2=-0.0025, 0, 0.02$.
Each figure contains the dynamical correlation function, scaling plot, and estimated temperature-dependent relaxation time $\tau$ for the order parameter of interest.
The system size used in the NER analyses was $L=1800$, and we confirmed that the system-size dependence is negligibly small.

First, we examine the results computed for $L=1800$ at $J_2=-0.0025$, which is away from the first-order transition line.
The results were averaged over 210 samples for different random seeds to mitigate the statistical fluctuations.
We discuss the results only for the octuple BKT transition because the transition temperatures of the magnetic BKT transitions and chiral transitions are too low.
In Fig.~\ref{fig:ner_j2_-2.5e-3_0}(a1), the time dependence of the dynamical correlation function becomes critical and decays algebraically below $T=0.078$.
To determine the critical temperature accurately, we executed scaling analyses using Eq.~\eqref{eq:scaling_low}.
The results are presented in Figs.~\ref{fig:ner_j2_-2.5e-3_0}(b1) and (c1).
The transition temperature was estimated as
\begin{align}
   T_{\mathrm{oct}} = 0.053 \pm 0.010,
\end{align}
where the error bar was estimated by dividing the 210 random samples into seven subgroups and computing the standard deviation of the results obtained for each subgroup.

We now turn to the results for $J_2=0$ computed with $L=1800$.
Figures~\ref{fig:ner_j2_-2.5e-3_0}(a2), (b2) and (c2) present results for \moctupole derived from 240 random samples. 
The time dependence of the dynamical correlation function was found to be similar to that for $J_2=-0.0025$.
Following the same procedure as before, we estimated the transition temperature to be
\begin{align}
   T_{\mathrm{oct}}= 0.071 \pm 0.005,
\end{align}
where we utilized six subgroups for estimating the error bar.
This result is consistent with the result of the previous study $T_{\mathrm{oct}}=0.070$--$0.076$~\cite{lee1986commensurate, Rzchowski1997, Korshunov2002, song2023tensor} within the error bar.

Finally, we examine the results for $J_2=0.02$ and $L=1800$ displayed in Fig.~\ref{fig:ner_j2_0.02}.
We found that 50 samples were enough for robust statistics because $J_2=0.02$ is away from the first-order transition.
The transition temperatures were estimated as follows:
\begin{align}
   T_{\sqrt{3}\times\sqrt{3}} &= 0.072  \pm 0.005,\\
   \Tafchiral                 &= 0.0701 \pm 0.0005,\\
   T_{\mathrm{oct}}           &= 0.096  \pm 0.003,
\end{align}
where we utilized five subgroups for estimating the error bars.
A noteworthy observation is that the transition temperature of the octupole QLRO, $T_{\mathrm{oct}}$, is well higher than $T_{\sqrt{3}\times\sqrt{3}}$ and $\Tafchiral$.
It is also noted that $T_{\sqrt{3}\times\sqrt{3}}$ and $\Tafchiral$ match within the error bars.
Furthermore, our estimation of the critical exponent yields $z\nu \simeq 6.3$, deviating from the expected value of $z\nu \simeq 2$~($\nu$=1 and $z\sim2$ for the two-dimensional Ising criticality~\cite{nishimori2010elements, Ito1993ner}).
This discrepancy may be attributed to the proximity effects of the antiferrochiral LRO and the \sqrtthreesqrtthree magnetic QLRO transitions.

\subsection{Region III ($J_2 \ge 4\times10^{-2}$)}\label{sec:regionIII}
In this subsection, we discuss the equilibrium MC results and the NER results for Region III.

\subsubsection{Results of equilibrium MC}
\begin{figure*}
   \begin{center}
  \includegraphics[width=0.8\linewidth]{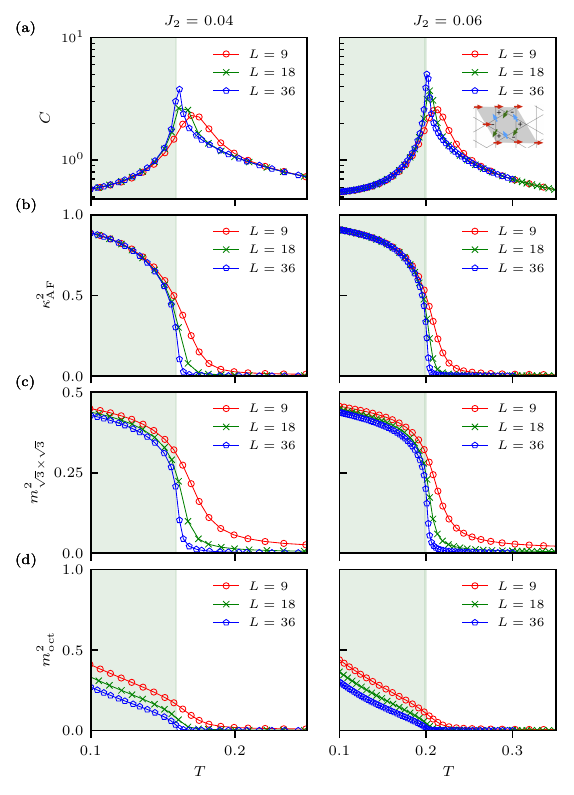}
   \end{center}
  \caption{Temperature dependence of the specific heat [(a)] and the order parameters \mqsqrtthree [(b)], \afchiral [(c)], \moctupole [(d)] computed at $J_2=0.04, 0.06$ in Region III.
           The system sizes were $L=9$, 18, and 36. 
           The vertical lines denote the transition temperature at the bulk limit estimated by NER.
  }
  \label{fig:eqvals_region3}
\end{figure*}

Figure~\ref{fig:eqvals_region3} displays the temperature dependence of the specific heat and the order parameters calculated for $L=9, 18, 36$ at various values of $J_2$.
Figure~\ref{fig:eqvals_region3}(a) presents the temperature dependence of the specific heat at $J_2=0.04$ and $0.06$.
For all the values of $J_2$, the specific heat exhibits a single peak.
For both $J_2=0.04$ and $0.06$, there is a slight increase in peak height with increasing $L$. 
Further analyses of the energy histogram support the continuous nature of the transition (not shown).

As depicted in Fig.~\ref{fig:eqvals_region3}(b), the system exhibits antiferrochiral LRO at low $T$.
The order parameters vanish continuously as $T$ increases, signifying the continuous nature of the transition.
The transition temperature appears to increase by increasing $J_{2}$.
As illustrated in Fig.~\ref{fig:eqvals_region3}(c), $\mqsqrtthree$ disappears similarly to $\afchiral$.
We will show that these two transition temperatures are close but separated by NER analyses.
Figure~\ref{fig:eqvals_region3}(d) plots the temperature dependence of $\moctupole$.
As previously mentioned, \moctupole becomes finite in the presence of the antiferrochiral LRO.
As $T$ increases, both \moctupole and \afchiral seem to disappear at the same temperature within the accuracy of the current simulations.
This observation indicates that the non-zero value of \moctupole originates from the primary \sqrtthreesqrtthree QLRO.

\subsubsection{Results of NER}
\begin{figure*}
   \begin{center}
   \includegraphics[width=0.9\textwidth]{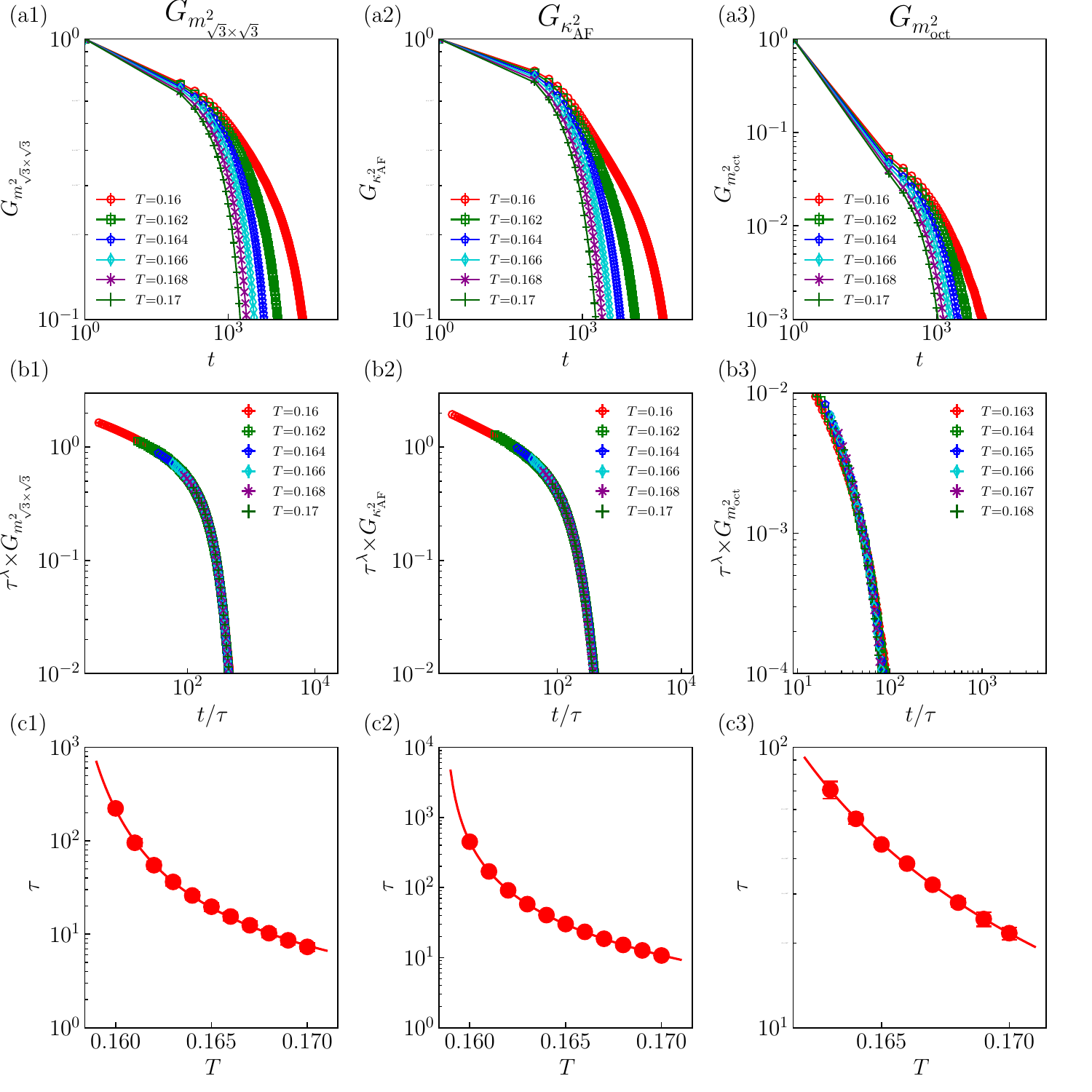}
   \caption{NER results for $J_2 = 0.04, L = 1800$. 
            The dynamical correlation function $G(t)$ is computed for the \sqrtthreesqrtthree order parameter \mqsqrtthree, the antiferrochiral order parameter \afchiral, and the octupole order parameter \moctupole.
            (a) $G(t)$ as a function of the MC step $t$.
            (b) Scaling plot for $G(t)$.
            (c) The relaxation time $\tau$ as a function of temperature $T$ in an arbitrary unit. 
                The curve represents a fit by BKT scaling $\tau = b \ \exp (c/\sqrt{T-T_{\mathrm{BKT}}})$ in (c1) and (c3) and by the power law $\tau = b (T-T_{\mathrm{c}})^{-z\nu}$ in (c2).
   }
   \label{fig:ner_j2_0.04}
   \end{center}
\end{figure*}

\begin{figure*}
   \begin{center}
   \includegraphics[width=0.9\textwidth]{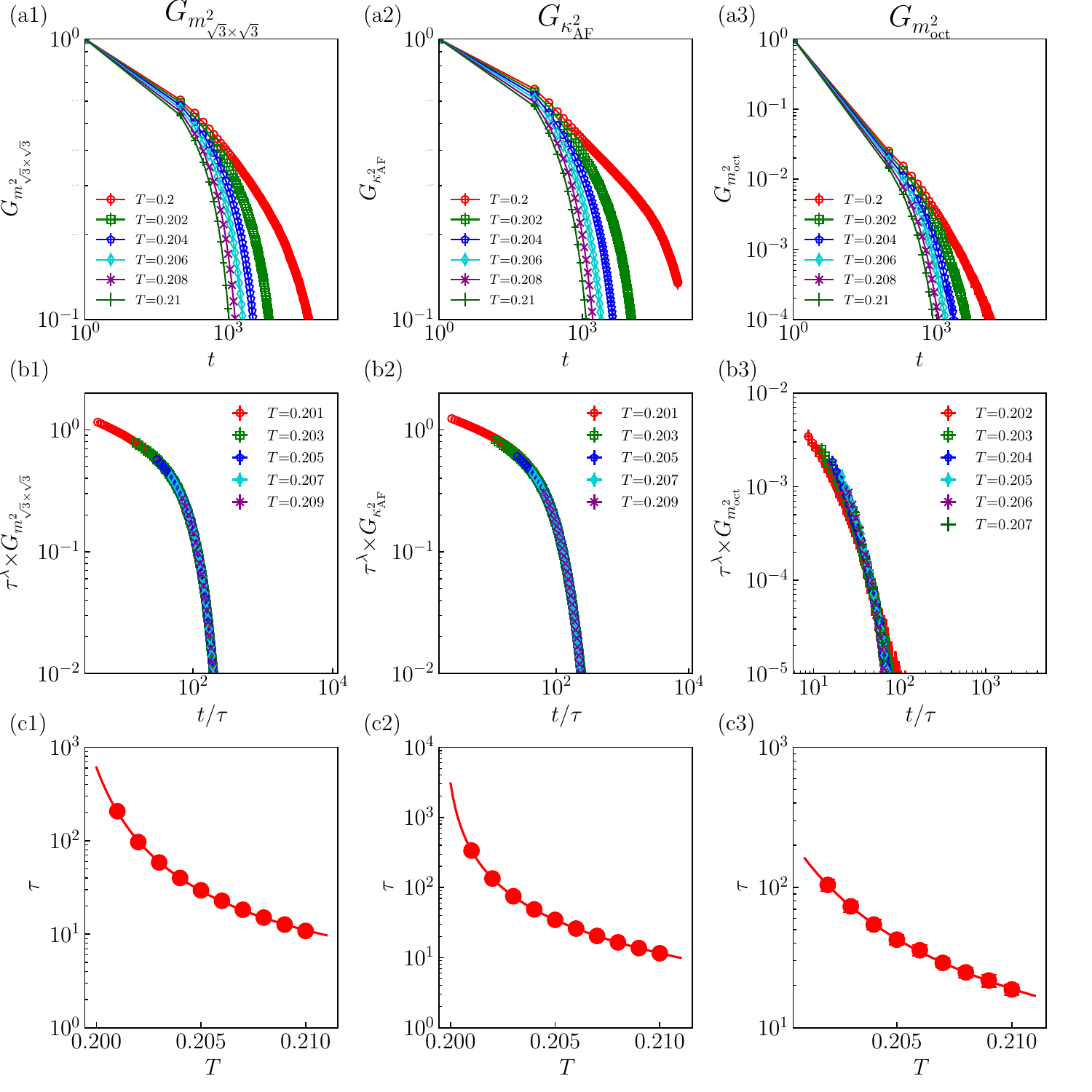}
   \caption{NER results for $J_2 = 0.06, L = 1800$. 
            The dynamical correlation function $G(t)$ is computed for the \sqrtthreesqrtthree order parameter \mqsqrtthree, the antiferrochiral order parameter \afchiral, and the octupole order parameter \moctupole.
            (a) $G(t)$ as a function of the MC step $t$.
            (b) Scaling plot for $G(t)$.
            (c) The relaxation time $\tau$ as a function of temperature $T$ in an arbitrary unit. 
                The curve represents a fit by BKT scaling $\tau = b \ \exp (c/\sqrt{T-T_{\mathrm{BKT}}})$ in (c1) and (c3) and by the power law $\tau = b (T-T_{\mathrm{c}})^{-z\nu}$ in (c2).
   }
   \label{fig:ner_j2_0.06}
   \end{center}
\end{figure*}

We now determine the transition temperatures using the NER method for $L=1800$.
Figures~\ref{fig:ner_j2_0.04} and ~\ref{fig:ner_j2_0.06} show the results for $J_2 = 0.04$ and 0.06, respectively.  
We utilized 50 random samples.
The transition temperatures for $J_2=0.04$ were estimated as follows:
\begin{align}
   T_{\sqrt{3}\times\sqrt{3}} &= 0.1562 \pm 0.0001, \\
   \Tafchiral                 &= 0.1590 \pm 0.0007,\\
   T_{\mathrm{oct}}      &= 0.151  \pm 0.004.
\end{align}

The transition temperatures for $J_2=0.06$ were estimated as follows:
\begin{align}
   T_{\sqrt{3}\times\sqrt{3}} &= 0.19735 \pm 0.00004,\\ 
   \Tafchiral                 &= 0.1995 \pm 0.0003,\\
   T_{\mathrm{oct}}      &=  0.195 \pm 0.001.
\end{align}
These results suggest that these three transitions nearly coincide, consistent with the MC results in Fig.~\ref{fig:eqvals_region3}.
As shown in Fig.~\ref{fig:ner_j2_0.06}(b3), the scaling collapse of the octupole ordering is relatively poor. 
The fast relaxation of $\moctupole$ may be the origin of this poor scaling collapse. 
However, we can obtain the smooth temperature dependence of $\tau$ and the estimated critical temperature seems to be reasonable.
Additionally, for both $J_2=0.04$ and $J_2=0.06$, our estimation of the critical exponent yields $z\nu \simeq 1.6$, which is closer to the expected value of $z\nu \simeq 2$~\cite{nishimori2010elements,Ito1993ner} than the result for $J_2=0.02$.

\section{Nature of the first-order transition}\label{sec:nature}
In this section, we discuss the nature of the first-order transition observed in Region I.
Specifically, we focus on the statistics of loops consisting of two types of spins formed during the loop update.

\begin{figure}[htb]
   \centering
   \includegraphics[width=0.8\columnwidth]{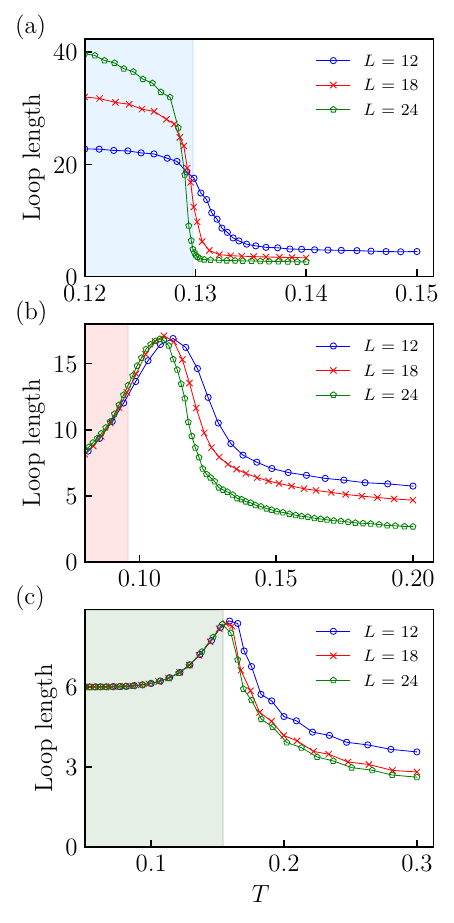}
   \caption{Average lengths of loops for $J_2=-0.02$ [(a)], $0.02$ [(b)], and $0.04$ [(c)]. 
            We plot the data for $L=12$, 18, and 24.
   }
   \label{fig:loop_length}
\end{figure}

Figure~\ref{fig:loop_length} illustrates the average loop lengths for typical values of $J_2$.
Firstly, we discuss the result for Region I [Fig.~\ref{fig:loop_length}(a)].
As highlighted in Sec.~\ref{sec:orders}, the loop lengths are $O(L)$ in the perfect $q=0$ LRO phase at $T=0$.
As anticipated, the average loop length increases approximately linearly with $L$ at low temperatures below the transition temperature.
Above the transition temperature ($T \simeq 0.13$), the average loop length diminishes as $L$ increases.
In other words, as $L$ escalates, the temperature dependence of the average loop length intensifies.
Such a precipitous change in the loop length from $O(L)$ to $O(1)$ necessitates a global alteration in spin configurations, leading to the appearance of a substantial energy barrier between the low-$T$ phase and the high-$T$ phase.
This might be the origin of the first-order transition.

Next, we proceed with the discussion of the results for $J_2=0.02$ [Fig~\ref{fig:loop_length}(b)].
In Region II, two successive transitions occur: one from the paramagnetic phase to the octupole QLRO phase and the other from the octupole QLRO phase to the coexisting phase of antiferrochiral LRO and $\sqrtthreesqrtthree$ QLRO.
As elucidated in Sec.~\ref{sec:orders}, the perfectly-ordered $\sqrtthreesqrtthree$ structure has the shortest loops of length 6.
The average loop length decreases below $T\simeq 0.09$ close to the octupole BKT transition temperature, which aligns with this zero-$T$ limit.
Additionally, the average loop length exhibits a broad peak near $T \simeq 0.12$ for all values of $L$.
The height of each peak is approximately 18 and is independent of $L$.
The emergence of the $L$-independent peak might be attributed to the existence of short localized loops generated by thermal fluctuations in the \sqrtthreesqrtthree spin configurations at low temperatures.

In Region III, as depicted in Fig.~\ref{fig:loop_length}(c), the average loop length converges to 6, as anticipated.

\section{Summary}\label{sec:summary}
In this study, we numerically investigated the cooperative effects of thermal fluctuations and next-nearest neighbor interactions $J_2$ on the macroscopically degenerate ground-state manifold of the classical $J_1$-$J_2$ $XY$ kagome antiferromagnet. 
We mapped out a $J_2$-$T$ phase diagram by extensive classical MC simulations using the equilibrium MC and the NER methods. 

Let us summarize our findings: (i)
We have discovered the first-order transition between the paramagnetic phase and the $q=0$ magnetic QLRO in the range $-0.07 < J_2 < -0.0075$.
We confirmed the first-order nature of the transition via an analysis of energy histograms and the finite-size scaling of the peak temperatures of the specific heat.
The first-order nature is most enhanced around $J_2=-0.02$ and weakens as $J_2$ approaches one of the endpoints of the transition.
(ii) We found that the octupole QLRO phase remains stable in the region $-0.005 < J_2 < 0.04$. 
We determined the $J_2$ dependence of the transition temperature precisely by the NER method.
However, at the small negative $J_{2}$ region, for example, $J_{2}=-0.005$, 
despite the specific heat displaying the low-$T$ peak at $T\simeq 0.06$,
our MC simulations were unable to confirm the existence of the ferrochiral LRO and the $q=0$ magnetic QLRO below this peak.
As illustrated in Fig.~\ref{fig:proposed_phase_diagram}, these orders might emerge at lower temperatures inaccessible by the current equilibrium MC or NER simulations.
(iii) For $J_2 \ge 0.04$, we precisely determined a $\sqrtthreesqrtthree$ BKT transition temperature and the antiferrochiral transition temperature.
(iv) We examined the origin of the first-order transition in the context of the average loop lengths for $-0.07 < J_2 < -0.0075$.

Before concluding this paper, we discuss potential future directions.
First, a similar first-order transition was reported in MC simulations for the classical $J_1$-$J_2$ Heisenberg antiferromagnet with antiferromagnetic $J_2$~\cite{spenke2012classical}.
An intriguing direction for future study would be to introduce an easy-axis anisotropy in the Heisenberg antiferromagnet, establishing a connection between the $XY$ and Heisenberg limits.
This could help shed light on the origin of the perplexing first-order transition in the Heisenberg limit.

Secondly, identifying the ferrochiral LRO and the $q=0$ magnetic QLRO at the small negative $J_2$ region, along with testing our proposed $J_2$-$T$ phase diagram (Fig.~\ref{fig:proposed_phase_diagram}), remains a challenging endeavor for future research.
This could help us better understand the low-temperature properties of the $J_1$-$J_2$ $XY$ kagome antiferromagnet for small antiferromagnetic $J_2$ as well as the nature of the first-order transition in the $J_1$-$J_2$ Heisenberg  antiferromagnet~\cite{spenke2012classical}.

\begin{acknowledgments}
F.K. and H.S. were supported by JSPS KAKENHI Grants No. 18H01158, No. 21H01041, and No. 21H01003, JST PRESTO Grant No. JPMJPR2012, Japan.
F.K. and H.S. thank T. Okubo for the fruitful discussions.
\end{acknowledgments}

\appendix

\section{Scaling analysis}\label{sec:scaling}

In this section, we explain the scaling analysis in detail. 
For $T \ge T_{\mathrm{BKT}}$, we assume the following scaling law
\begin{align}
   g(t/\tau) = \tau^{\lambda}(T)G(t, T),
   \label{eq:scaling_low}
\end{align}
where $G(t, T)$ represents the temperature-dependent dynamical correlation function at time $t$ measured in units of MC steps, $g(x)$ is a temperature-independent function, and $\lambda$ denotes the dynamical critical exponent of $G(t, T)$. 
We optimize the cost function defined as follows:
\begin{align}
   F[\vec{\tau}, \lambda] \equiv \frac{\sum_{i=1}^{M}\sum_{j=1}^{N_T}|\ln\qty[\tau^{\lambda}(T_j)G(t_i, T_j)] - \tilde{g}(t_i/\tau)|^2} {\sum_{k=1}^{M}|\tilde{g}(t_k/\tau)|^2},
   \label{eq:cost_func}
\end{align}
where $\vec{\tau}$ represents the vector of $N_T$ values of $\tau$ and $\tilde{g}(t/\tau)$ is defined as $\tilde{g}(t/\tau) \equiv \frac{1}{N_T} \sum_{i=1}^{N_T} \ln\qty[\tau^{\lambda}(T_i)G(t, T_i)]$. 
We now introduce $N_T$ as the number of temperatures and $M$ as the number of sampling points of time $t$.
To compute $\tilde{g}(t/\tau)$, we interpolate $G(t, T)$ along the $t$ axis using linear interpolation. 
We use the Nelder-Mead method~\cite{Nelder1965simplex} to optimize the cost function~\eqref{eq:cost_func}.

\bibliography{main}

\end{document}